\documentclass[conference]{IEEEtran}
\usepackage{fancyhdr}
\IEEEoverridecommandlockouts
\usepackage{cite}
\usepackage{amsmath,amssymb,amsfonts}
\usepackage{algorithmic}
\usepackage{graphicx}
\usepackage{textcomp}
\usepackage{xcolor}
\def\BibTeX{{\rm B\kern-.05em{\sc i\kern-.025em b}\kern-.08em
    T\kern-.1667em\lower.7ex\hbox{E}\kern-.125emX}}

\usepackage{amsmath, amsthm}
    
\usepackage{graphicx, xcolor} 
\usepackage{hyperref, url}
\usepackage{threeparttable}
\usepackage{array}
\usepackage{amsthm}

\usepackage{booktabs, multirow}
\usepackage{caption}
\usepackage{subcaption}
\usepackage{enumitem}
\usepackage{setspace}

\usepackage[most]{tcolorbox}
\tcbuselibrary{theorems}
\usepackage[most]{tcolorbox}
\usepackage{amsmath,amsthm}
\usepackage{xparse} 

\NewDocumentEnvironment{boxedlemma}{O{}}
  {\begin{tcolorbox}[colback=gray!5,
                     colframe=black,
                     boxrule=0.5pt,
                     arc=1mm,
                     top=1mm, bottom=1mm, left=1mm, right=1mm,
                     enhanced]
   \refstepcounter{lem}
   \textbf{Lemma \thelem.} \label{#1}%
   \ignorespaces}
  {\end{tcolorbox}}

\usepackage[ruled, linesnumbered, norelsize]{algorithm2e}
\usepackage{colortbl}
\definecolor{dg}{HTML}{C8E6C9}  
\definecolor{lg}{HTML}{E8F5E9}  
\definecolor{ly}{HTML}{FFF9C4}  
\definecolor{lr}{HTML}{FFCDD2}  
\definecolor{dr}{HTML}{EF9A9A}  

\begin{document}

\title{TOPIQ: Statistical Error Propagation for Quantity-of-Interest Prediction under Lossy Compression
}

 \author{\IEEEauthorblockN{Youyuan Liu\IEEEauthorrefmark{1}, Bo Jiang\IEEEauthorrefmark{1}, Taolue Yang\IEEEauthorrefmark{1}, Sheng Di\IEEEauthorrefmark{2}, Robert Underwood\IEEEauthorrefmark{2}, Sian Jin\IEEEauthorrefmark{1}}
 \IEEEauthorblockA{\IEEEauthorrefmark{1}\textit{Department of Computer \& Information Sciences},
 \textit{Temple University},
 Philadelphia, USA}
 \IEEEauthorblockA{\IEEEauthorrefmark{2}\textit{Mathematics and Computer Science Division},
 \textit{Argonne National Laboratory},
 Lemont, USA}
 }

\maketitle

\thispagestyle{fancy}
\lhead{}
\rhead{}
\chead{}
\lfoot{\footnotesize{SC26, November 15-20, 2026, Chicago, Illinois, USA
\newline 979-8-3195-4789-7/26/\$31.00 \copyright 2026 IEEE}}
\rfoot{}
\cfoot{}
\renewcommand{\headrulewidth}{0pt}
\renewcommand{\footrulewidth}{0pt}

\begin{abstract}
Lossy compression is essential for managing massive scientific data, but per-element error bounds do not translate into bounds on downstream quantities of interest (QoIs) such as regional averages, neural network predictions, or multi-field derived quantities. We present TOPIQ, a statistical error-propagation framework that predicts QoI-level bias and uncertainty from compact compression metadata (less than 0.1\% of original data). TOPIQ decomposes QoIs into primitive operators with closed-form propagation rules accounting for spatial error correlation and data-error coupling; new QoIs are supported by composition at runtime with no per-QoI derivation or retraining. Across 552 evaluations spanning 4 datasets, 3 compressors, 4 QoI families, and 8 error bounds, 93.1\% of configurations achieve well-calibrated predictions. Pre-computed metadata enables post-hoc uncertainty quantification for arbitrary query regions at 56x--402x speedup over direct computation. A case study demonstrates integration into an AI-driven analysis pipeline with end-to-end confidence intervals for dynamically composed queries.
\end{abstract}


\section{Introduction}

Large-scale scientific simulations produce data at rates that far exceed storage and I/O capacity.
Climate simulations can generate hundreds of terabytes every 16 seconds~\cite{liu2025vldb,jiao2022vldb}; Cosmology simulation such as Nyx simulation at $4096^3$ resolution produces 2.8\,TB per snapshot, reaching petabyte scale over a full campaign~\cite{nyx}; and fusion applications on exascale systems are projected to generate exabytes per run~\cite{liu2025vldb,jiao2022vldb}.
This data volume poses fundamental challenges for storage, transfer, and post-hoc analysis, making compression a practical necessity rather than an optimization.
Since lossless compression offers limited reduction for floating-point scientific data, error-bounded lossy compression has become the standard approach, offering substantially higher compression ratios while controlling per-element reconstruction error~\cite{sz3,zfp,SPERR,yang2026dcc}.

However, data-level error bounds do not directly translate into guarantees on downstream \emph{quantities of interest} (QoIs), the derived analysis results that ultimately drive scientific interpretation and decision making.
Scientists rarely inspect raw field values; they compute regional averages, derived physical quantities, or run inference models over compressed data.
The gap between data-level fidelity and QoI-level reliability is especially critical as scientific workflows become more dynamic: AI-driven analysis pipelines and autonomous agents increasingly query compressed datasets at arbitrary locations and with QoIs not known at compression time~\cite{liu2025beyond,jiang2025geogridbench}, making it impractical to fix the analysis target before compression.
What is needed is a way to reason about the effect of compression on a broad class of QoIs, at any query location, after the data has already been compressed.

Compression errors propagate through application workflows in a QoI-dependent manner. Some QoIs remain stable under substantial perturbation, whereas others, especially those involving spatial aggregation, may amplify pixel-level errors by orders of magnitude when spatial correlation is present in the error field~\cite{liu2025vldb}.
Traditional data-level metrics (pointwise error, PSNR, SSIM) are insufficient to capture these effects, because the relationship between data distortion and QoI distortion depends on the QoI's computational structure, the spatial correlation of errors, and the coupling between data values and their compression errors~\cite{qiu2025scworkshop}.

Existing work has mainly addressed this challenge through QoI-preserving compression. MGARD~\cite{ainsworth2019mgard} provides guaranteed error control for bounded linear QoIs. cpSZ~\cite{liang2020cpsz} and its extensions~\cite{toposz,liang2022tvcg,xia2024icde} preserve geometric features such as critical points and contour trees by deriving sufficient pointwise error bounds from the target QoI. QPET~\cite{liu2025vldb} further extends this idea to differentiable symbolic QoIs, primarily in the univariate setting, with limited multivariate support such as vector magnitude, through portable numerical error-bound tuning. Related efforts have also explored compressor tuning toward standard quality metrics, storage-budget-driven error-bound estimation, and learned quality prediction~\cite{liu2022sc,rahman2023icde,mumenin2024qualitynet,deepcq2025}, but these do not model how compression errors propagate through downstream QoIs. All of the above are most effective when the target QoI is specified at compression time; their support is much weaker when the QoI is not known in advance or when analysis is performed later on previously compressed regions.

This limitation is becoming increasingly pressing as LLM-driven agents and autonomous analysis pipelines are deployed over scientific data~\cite{xie2025wildfiregpttailoredlargelanguage,li2023autonomousgisnextgenerationaipowered,climategpt}. Such systems dynamically compose queries at runtime—targeting arbitrary spatial regions, derived quantities, or multi-field comparisons making it impractical to anticipate every QoI at compression time.

We propose TOPIQ, a statistical error-propagation framework that addresses these limitations.
Rather than preserving QoIs during compression or predicting quality metrics, TOPIQ characterizes the effect of compression on \emph{any runtime-defined QoI expressible through supported smooth primitive operators} as a calibrated confidence interval, bias $\pm$ uncertainty, derived from compact error metadata ($<0.1\%$ of the original data size).
TOPIQ decomposes QoIs into primitive operators (linear, nonlinear, aggregation), each with a closed-form propagation rule that accounts for spatial error correlation and data--error coupling.
New QoI are supported by composition at runtime, with no per-QoI derivation or retraining, a property that directly enables integration into AI-driven workflows where the QoI is constructed dynamically from user queries.

Our main contributions are:
\begin{enumerate}[nosep,leftmargin=*]
  \item \textbf{A statistical error-propagation framework} that predicts QoI-level bias and variance from compact per-block metadata, supporting runtime compositions of smooth linear, nonlinear, and aggregation operators.
  New QoIs are handled by composition, with no per-QoI derivation or retraining.

  \item \textbf{Comprehensive empirical validation} across 4 scientific datasets, 3 compressors (SZ3, SPERR, ZFP), and 4 QoI families including neural networks and multi-field ratios.
  Across 552 evaluations (several fields from 4 datasets, 3 compressors, 4 QoI families, 8 error bounds), 93.1\% achieve well-calibrated predictions ($\sigma_z \in [0.7, 1.3]$); the failures are concentrated in two diagnosable mechanisms.

  \item \textbf{Post-hoc spatial queries} via pre-computed metadata ($<0.1\%$ of original data), enabling uncertainty quantification for arbitrary query regions without retaining the original data, and $56{\times}$ to $402{\times}$ faster than direct error computation on CESM-ATM dataset~\cite{sdrbench}.

  \item \textbf{A case study} demonstrating integration into an AI-driven scientific analysis pipeline, where TOPIQ provides end-to-end confidence intervals for dynamically composed queries, a capability that QoIs-preserving compressors cannot offer for off-grid or runtime-defined QoIs.
\end{enumerate}

\section{Background}

\subsection{Error-Bounded Lossy Compression}
\label{sec:lossy-compression}

Lossy compression is one of the most effective techniques for reducing the massive data volumes generated by scientific simulations.
Unlike lossless compression, which preserves data exactly but offers limited reduction on floating-point arrays, lossy compression achieves significantly higher compression ratios by allowing controlled information loss.
A new generation of lossy compressors, including prediction based compressors, such as SZ~\cite{sz3, sz3algo,sz1.4,sz2}, and transform based compressors such as ZFP~\cite{zfp}, SPERR~\cite{SPERR}, and MGARD~\cite{ainsworth2019mgard,mgardplus}, has been developed specifically for scientific floating-point data and is now widely integrated into HPC systems to reduce storage demands and accelerate I/O~\cite{jin2022sc,jin2023eurosys}.

A key distinction between scientific lossy compressors and general-purpose codecs (e.g., JPEG~\cite{taubman2012jpeg2000}) is the ability to provide strict, user-specified error controls.
In \emph{error-bounded} mode, the user specifies an error type (absolute, relative, or pointwise relative) and a bound (e.g., $\varepsilon = 10^{-3}$); the compressor then guarantees that every reconstructed value stays within that bound.
However, these metrics measure aggregate distortion of the reconstructed field and do not characterize how errors affect specific downstream computations, a gap that motivates the QoI-level analysis in this work.

\subsection{Quantities of Interest}
\label{sec:qoi-background}

Scientists rarely inspect raw field values; they compute derived results that drive analysis and decision making.
We refer to these derived results as \emph{quantities of interest} (QoIs).

QoIs span a wide complexity spectrum.
\emph{Pointwise transforms} ($x^2$, $\sin x$, $e^x$) operate on individual values.
\emph{Regional aggregates} (block means, weighted sums) reduce a spatial neighborhood to a single statistic.
\emph{Multivariate} QoIs combine multiple fields, from simple expressions like vector magnitude $\sqrt{v_x^2+v_y^2+v_z^2}$ to complex compositions such as the \emph{cloudy radiative ratio} used in climate analysis~\cite{ramanathan1989cre,zelinka2012kernels},which is shown in Section ~\ref{sec:composing-qoi}.

Lossy compression introduces a fundamental challenge for QoI reliability.
An error bound on individual data values does \emph{not} translate into a bound on the QoI: compression errors can be amplified, attenuated, or systematically biased as they propagate through the QoI computation, depending on the QoI's structure, the spatial correlation of the errors, and the coupling between data values and their errors.
Traditional metrics such as PSNR and SSIM do not capture these QoI-dependent effects.
Understanding and predicting how compression errors propagate through QoIs is therefore essential for scientists to trust compressed data in their analysis workflows.

\begin{figure}[t]
\centering
\includegraphics[width=0.95\linewidth]{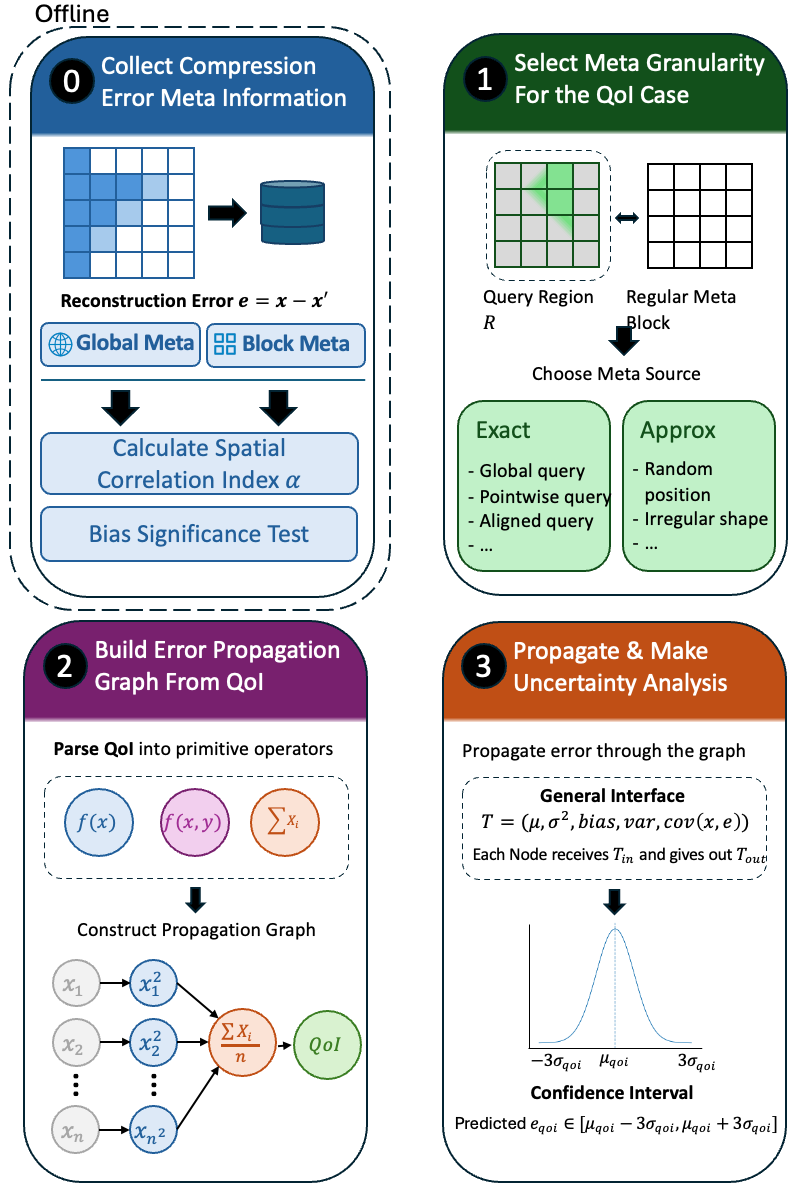}
\caption{Overview of TOPIQ. The framework decouples one-time offline compression-error metadata extraction from online query-time QoI uncertainty analysis. Offline, TOPIQ collects global and block-level error metadata and calibrates spatial correlation and bias significance. Online, it selects the appropriate metadata granularity for the queried QoI, constructs an error-propagation graph from the QoI definition, and propagates the error tuple through the graph to produce the final confidence interval.}
\vspace{-1em}
\label{fig:overview}
\end{figure}

\subsection{Predicting Compression Impact in Scientific Workflows}

Existing approaches address this question at different levels. At the compression-ratio level, FXRZ~\cite{rahman2023icde} predicts the error bound needed to meet a storage budget, but does not characterize downstream analysis impact. At the data-quality level, QoZ~\cite{liu2022sc} auto-tunes compressors toward metrics such as PSNR, and learned surrogates like QualityNet and DeepCQ~\cite{mumenin2024qualitynet,deepcq2025} predict quality metrics from data features and compression settings. However, metrics like PSNR do not capture QoI-level effects such as error amplification through spatial aggregation, and existing surrogates target predefined metrics with per-dataset training. None of these methods reason about how compression errors propagate through a downstream QoI. 
TOPIQ fills this gap: given compact error metadata, it predicts the bias and uncertainty of \emph{any QoI expressible as a composition of supported smooth operators}, enabling practitioners to assess whether a compression configuration is safe for their specific analysis before committing to it.

\subsection{QoI-Preserving Lossy Compression}

QoI-preserving lossy compression guarantees that the error in a specified QoI stays within a user-given tolerance, in addition to the standard data-level error bound.
This is typically achieved by deriving sufficient pointwise error bounds from the target QoI and using them to guide per-pixel compression.

MGARD~\cite{ainsworth2019mgard} is among the first compressors to enable QoI preservation, providing guaranteed error control for bounded linear QoIs via multilevel decomposition.
cpSZ~\cite{liang2020cpsz} extends this to geometric features: it derives pointwise error bounds that preserve critical points in piecewise-linear vector fields, and subsequent work covers multilinear fields, contour trees, and other topological features.
QPET~\cite{liu2025vldb} further generalizes to differentiable symbolic QoIs, primarily univariate (e.g., $x^2$, $\sin x$, $e^x$), with limited multivariate support, through a portable numerical layer that works across compressors.

These methods share two structural constraints: (1)~the QoI and analysis region must be fixed at compression time, and (2)~extending to a new QoI requires case-specific derivation of pointwise error bounds.
This makes them well-suited for fixed analysis pipelines but incompatible with dynamic, exploratory workflows where the QoI is defined at query time.
TOPIQ does not replace these methods; rather, it complements them.
When a deterministic guarantee on a specific QoI is required and the target is known at compression time, QoI-preserving compressors (MGARD for bounded linear functionals, cpSZ for topological features, QPET for differentiable symbolic functions) are the appropriate choice.
When the QoI is defined post-hoc, composed at runtime, or targets regions not fixed during compression, TOPIQ provides calibrated statistical bounds from pre-computed metadata.
Table~\ref{tab:qoi_lc_landscape} summarizes these directions and positions TOPIQ.

\begin{table*}[t]
\caption{Comparison of application-aware lossy compression approaches.
``Ex Post Facto QoI'' indicates whether the method supports QoIs that are defined after compression (e.g., by an autonomous agent or user query), as opposed to requiring the QoI to be specified before or during compression.
TOPIQ is the only approach that provides uncertainty estimates for runtime-defined QoIs.}
\label{tab:qoi_lc_landscape}
\centering
\footnotesize
\renewcommand{\arraystretch}{1.15}
\setlength{\tabcolsep}{3pt}
\begin{tabular}{p{3cm} p{3.5cm} p{4.8cm} p{4.0cm} p{1.6cm}}
\toprule
\textbf{Approach} & \textbf{Repr.} & \textbf{Strengths} & \textbf{Limitations} & \textbf{Ex Post Facto QoI} \\
\midrule
Quality-metric tuning
  & QoZ~\cite{liu2022sc}
  & High CR for PSNR etc.
  & Quality metrics, not QoIs; tied to compressor
  & No \\
Fixed-ratio control
  & FXRZ~\cite{rahman2023icde}
  & Meets storage budgets
  & Not QoI-oriented
  & No \\
ML quality surrogate
  & QualityNet,DeepCQ~\cite{mumenin2024qualitynet,deepcq2025}
  & Fast screening; cross-dataset
  & Training cost; predefined metrics only
  & No \\
QoI preservation
  & MGARD,cpSZ,QPET~\cite{ainsworth2019mgard,liang2020cpsz,liu2025vldb}
  & Deterministic guarantee; portable (QPET)
  & QoI \& region fixed at compression; case-specific derivation
  & No \\
\midrule
\textbf{Statistical propagation}
  & \textbf{This work}
  & \textbf{Composable; no training; tiny metadata}
  & \textbf{Statistical, not deterministic}
  & \textbf{Yes} \\
\bottomrule
\end{tabular}
\end{table*}

\section{Methodology}

Rather than deriving a new error analysis for each downstream metric, TOPIQ views a quantity of interest as a computation graph of primitive operators and propagates compact error statistics through this graph to predict the QoI-level bias and uncertainty.
This section first defines the error tuple that serves as the universal interface between operators (\S\ref{sec:tuple}), then derives the propagation rules for three operator families: single-field (\S\ref{sec:singlefield}), multi-field (\S\ref{sec:multifield}), and aggregation (\S\ref{sec:aggregation}).
Figure~\ref{fig:overview} illustrates the overall workflow, which separates a one-time offline metadata extraction phase from the online query-time propagation.

\subsection{Notation and Error Tuple}
\label{sec:tuple}

TOPIQ propagates uncertainty by passing a compact \emph{error tuple} through each operator in the computation graph.
Every operator takes one or two tuples as input and produces one as output; the tuple serves as the universal interface throughout the computation graph.

We define the five-tuple $T = (\mu, \sigma^2, b, v, c)$ whose components are listed in Table~\ref{tab:symbols}.

\begin{table}[t]
\centering
\caption{Symbol definitions.  The tuple $T = (\mu, \sigma^2, b, v, c)$ is the input and output of every operator; $v_u$, $v_c$, and $\alpha$ are introduced in \S\ref{sec:aggregation} for aggregation.}
\label{tab:symbols}
\small
\renewcommand{\arraystretch}{1.1}
\begin{tabular}{@{}cl@{}}
\toprule
Symbol & Definition \\
\midrule
$x$ & Original (uncompressed) data value \\
$\hat{x} = x + e$ & Reconstructed (decompressed) value \\
$e = \hat{x} - x$ & Compression error \\
\midrule
\multicolumn{2}{@{}l}{\emph{Error tuple $T = (\mu, \sigma^2, b, v, c)$}} \\
$\mu = \mathbb{E}[x]$ & Mean of the original data \\
$\sigma^2 = \mathrm{Var}(x)$ & Variance of the original data \\
$b = \mathbb{E}[e]$ & Bias (mean) of the compression error \\
$v = \mathrm{Var}(e)$ & Variance of the compression error \\
$c = \mathrm{Cov}(x, e)$ & Within-field data-error covariance \\
\midrule
\multicolumn{2}{@{}l}{\emph{Aggregation-specific (\S\ref{sec:aggregation})}} \\
$n$ & Number of pixels in a spatial block \\
$\delta_i$ & Per-pixel output error after operator(s) \\
$\bar{c}$ & Mean pairwise covariance among errors in a block \\
$v_u$ & Uncorrelated (pixel-level) component of $v$ \\
$v_c$ & Correlated (common-mode) component of $v$ \\
$\alpha$ & Spatial correlation index \\
$r = v_c / v$ & Correlation ratio (approximately scale-invariant) \\
\bottomrule
\end{tabular}
\vspace{-1em}
\end{table}

The pair $(\mu, \sigma^2)$ describes the signal, while the triple $(b, v, c)$ characterizes the compression error and its coupling with the data.
Intuitively, the reconstructed value~$\hat{x}$ has mean $\mu + b$ and variance $\sigma^2 + v + 2c$; the tuple keeps signal and error statistics cleanly separated so that both remain identifiable after extended operator chains.

In general, error-bounded scientific compressors are designed to be unbiased, and the per-block mean error is indeed negligibly small in most cases.
However, certain data/compressor combinations do produce a small but genuine systematic bias.
TOPIQ always propagates the measured global bias $b = \overline{e}$ (mean compression error across all pixels), because it represents the deterministic offset remaining after spatially varying block-level fluctuations have been captured by the $v_u/v_c$ decomposition and the $\alpha$ model.
Per-block mean errors, by contrast, overlap with the correlated variance already modeled by $v_c$; propagating the global mean avoids this double-counting.

\subsection{Single-Field Operators}
\label{sec:singlefield}

When an operator $f$ acts on a single compressed field, every reference to that field carries the \emph{same} compression error at each pixel, not an independent copy.
The cross-field independence assumption central to multi-field operators (\S\ref{sec:multifield}) therefore does not apply: in a product like $x^2 = x \cdot x$, both factors share identical errors, so the cross terms are nonzero and cannot be dropped.
This is why single-field and multi-field operators require different propagation rules.
Concretely, treating $x^2$ as a cross-field product would assume two independent error sources and underestimate the output variance (compare Eq.~\ref{eq:square-tuple} with Eq.~\ref{eq:mul-var}).

We further split single-field operators into linear and nonlinear.
Linear operators (constant addition, scalar multiplication) propagate exactly through the tuple with no approximation.
Nonlinear operators (e.g., $x^2$, $\sigma(x)$, $1/x$) require a second-order Taylor expansion.

\subsubsection{Linear: Constant Addition and Scalar Multiplication}

Linear operators propagate exactly through the tuple.
Adding a constant $a$ shifts only the mean:
\vspace{-4pt}
\begin{equation}
T_{x+a} = (\mu + a,\; \sigma^2,\; b,\; v,\; c)
\vspace{-4pt}
\end{equation}

Scalar multiplication $f(x) = kx$ scales each component by the appropriate power of $k$:
\vspace{-4pt}
\begin{equation}
T_{kx} = (k\mu,\; k^2\sigma^2,\; kb,\; k^2 v,\; kc)
\label{eq:linear-tuple}
\vspace{-4pt}
\end{equation}

The output is itself a five-tuple with the same structure, ready to be consumed by the next operator in the graph. And this is also the shared interface between all kinds of operators.

\subsubsection{Nonlinear: Second-Order Expansion}

For a general differentiable function $f$, linear propagation no longer holds; we approximate the output statistics through a second-order Taylor expansion~\cite{deltamethod}.

Let $\hat{x} = x + e$.
Expanding $f(\hat{x})$ around $\mu = \mathbb{E}[x]$, with $\hat{x} = \mu + (x - \mu) + e$:
\begin{equation}
f(\hat{x}) \approx f(\mu) + f'(\mu)\bigl[(x{-}\mu) + e\bigr] + \tfrac{1}{2} f''(\mu)\bigl[(x{-}\mu) + e\bigr]^2
\label{eq:delta-expand}
\end{equation}

Taking expectations and separating data variability from error yields the output tuple $T_f = (\mu_f, \sigma^2_f, b_f, v_f, c_f)$:
\begin{align}
\mu_f &= f(\mu) + \tfrac{1}{2} f''(\mu) \cdot \sigma^2, \label{eq:delta-mu} \\
\sigma^2_f &= \bigl[f'(\mu)\bigr]^2 \cdot \sigma^2 \label{eq:delta-sig} \\
b_f &= f'(\mu) \cdot b + \tfrac{1}{2} f''(\mu) \cdot v + f''(\mu) \cdot c \label{eq:delta-bias} \\
v_f &= \bigl[f'(\mu)\bigr]^2 \cdot v \label{eq:delta-var} \\
c_f &= f'(\mu) \cdot c \label{eq:delta-cov}
\end{align}

Equation~\eqref{eq:delta-bias} reveals three bias sources: first-order propagation of input bias ($f' \cdot b$), a Jensen-inequality correction from error variance ($\tfrac{1}{2} f'' \cdot v$), and a correction from data-error coupling ($f'' \cdot c$).
The output $T_f$ has the same five-component structure as the input and can be passed directly to the next operator.

\paragraph{Example: $f(x) = x^2$.}
With $f'(x) = 2x$ and $f''(x) = 2$, the per-pixel output error is $\delta_i \approx 2x_i \cdot e_i$.
Since the derivative $2x_i$ varies across pixels, the variance averages over the pixel population rather than evaluating at the mean alone:
\begin{equation}
v_{x^2} = 4\,\mathbb{E}[x^2] \cdot v = 4(\mu^2 + \sigma^2)\,v
\label{eq:square-var}
\end{equation}
using $\mathbb{E}[x^2] = \mu^2 + \sigma^2$.
The complete output tuple is:
\begin{equation}
T_{x^2} = \bigl(\mu^2 {+} \sigma^2,\;\; 4\mu^2 \sigma^2,\;\; 2\mu b {+} v {+} 2c,\;\; 4(\mu^2{+}\sigma^2)v,\;\; 2\mu c\bigr)
\label{eq:square-tuple}
\end{equation}
If $x^2$ were instead treated as a cross-field product of two independent copies, the variance would be $2(\mu^2{+}\sigma^2) v$ (Eq.~\ref{eq:mul-var}), exactly half the correct $4(\mu^2{+}\sigma^2) v$: the shared-error cross term doubles the variance, which the independent-copy model cannot capture.

\subsection{Multi-Field Operators and Cross-Field Independence}
\label{sec:multifield}

When an operator combines two separately compressed fields $x$ and $w$, cross-field covariance terms arise.
Unlike the single-field case where the same error appears on both sides of $x^2$, multi-field operators involve two distinct error sources $e_x$ and $e_w$ from independent compression runs.

\begin{figure}[t]
\centering
\begin{subfigure}{0.48\columnwidth}
  \includegraphics[width=\linewidth]{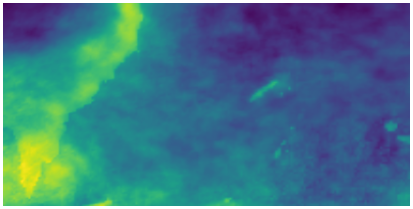}
  \caption{Original CLDTOT}
\end{subfigure}\hfill
\begin{subfigure}{0.48\columnwidth}
  \includegraphics[width=\linewidth]{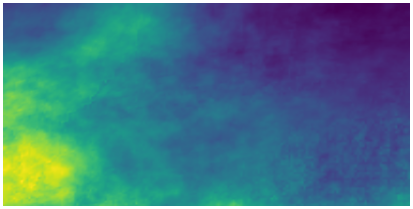}
  \caption{Original CLDHGH}
\end{subfigure}\\[3pt]
\begin{subfigure}{0.48\columnwidth}
  \includegraphics[width=\linewidth]{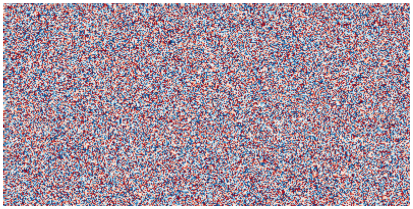}
  \caption{Error CLDTOT}
\end{subfigure}\hfill
\begin{subfigure}{0.48\columnwidth}
  \includegraphics[width=\linewidth]{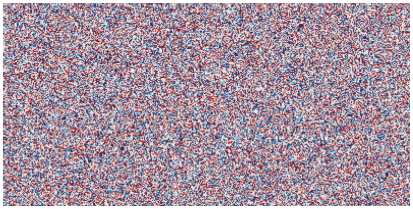}
  \caption{Error CLDHGH}
\end{subfigure}
\caption{Cross-field independence on CESM-ATM (SZ3, $\varepsilon_{\mathrm{rel}}{=}10^{-3}$).
The original fields (a, b) share visible spatial structure (Pearson $r = 0.43$).
Their compression errors (c, d) are nearly uncorrelated ($r = -0.002$), supporting the cross-field independence assumption.}
\label{fig:cross-field-corr}
\vspace{-1em}
\end{figure}

Since each field is compressed independently, the compression errors of different fields share no generative mechanism: distinct prediction contexts, transform coefficients, and quantization decisions.
The original data fields may be physically correlated (e.g., temperature and pressure), but their errors exhibit substantially weaker cross-field correlation than the data themselves (Figure~\ref{fig:cross-field-corr}).
We therefore adopt a \textbf{cross-field independence assumption}, treating $e_x$ and $(w, e_w)$ as independent.
The within-field data-error covariance $c = \mathrm{Cov}(x, e_x)$, which drives nonlinear bias (Eq.~\ref{eq:delta-bias}), remains fully tracked in the tuple; what we drop is only the cross-field coupling, whose impact is modest relative to the dominant $\mu^2 v$ terms, as confirmed by the ablation study in \S\ref{sec:ablation}.

Under this assumption, both linear and nonlinear cross-field operations follow the same combination rules, so no linear/nonlinear distinction is needed: the key question is independence between error sources, not the form of the operator.
This also eliminates $O(k^2)$ pairwise covariance tracking across $k$ fields, preserving composability as the graph deepens.

Existing QoI-preserving compressors provide only limited multi-field support.
TOPIQ's compositional approach extends to general multi-field QoIs built from the supported operators, such as the cloudy ratio in Figure~\ref{fig:qoi-graphs}(b), by composing the same primitives.

\subsubsection{Addition and Subtraction}

For $f(x, w) = x \pm w$ with two independently compressed fields, the output tuple is:
\vspace{-4pt}
\begin{equation}
T_{x \pm w} = \bigl(\mu_x \pm \mu_w,\; \sigma^2_x + \sigma^2_w,\; b_x \pm b_w,\; v_x + v_w,\; c_x \pm c_w\bigr)
\label{eq:add-two-field}
\vspace{-4pt}
\end{equation}

The variances $\sigma^2$ and $v$ always add regardless of the sign, since the two fields' errors are uncorrelated under the independence assumption.

\subsubsection{Multiplication}
\label{sec:multiplication}

Multiplication $f(x, w) = x \cdot w$ is the most important binary operation.
Let $\delta = \hat{x}\hat{w} - xw$ be the output error.
Expanding:
\vspace{-4pt}
\begin{equation}
\delta = \underbrace{w \cdot e_x + x \cdot e_w}_{\text{leading terms}} + \underbrace{e_x \cdot e_w}_{\approx\,0}
\label{eq:mul-expand}
\vspace{-4pt}
\end{equation}

The variance of $\delta$ has within-field and cross-field groups:
\begin{equation}
\mathrm{Var}(\delta) =
\underbrace{\mathrm{Var}(w e_x) {+} \mathrm{Var}(x e_w)}_{\text{(i) within-field, retained}}
{+}\, \underbrace{2\,\mathrm{Cov}(we_x,\, xe_w)}_{\text{(ii) cross-field, dropped}}
{+}\, \cdots
\label{eq:mul-var-full}
\end{equation}

Group~(i) requires only within-field quantities ($c_x$, $c_w$), which the tuple already carries.
Group~(ii) requires cross-field covariances ($\mathrm{Cov}(x, e_w)$, $\mathrm{Cov}(e_x, e_w)$, $\mathrm{Cov}(x, w)$), which we drop under the independence assumption.
The resulting output tuple takes the form:
\vspace{-4pt}
\begin{equation}
T_{xw} = \bigl(\mu_x \mu_w,\;\; \mu_x^2 \sigma^2_w {+} \mu_w^2 \sigma^2_x {+} \sigma^2_x \sigma^2_w,\;\; b_{xw},\;\; v_{xw},\;\; c_{xw}\bigr)
\vspace{-4pt}
\end{equation}
where:
\vspace{-4pt}
\begin{align}
b_{xw} &= \mu_w \cdot b_x + \mu_x \cdot b_w \label{eq:mul-bias} \\
v_{xw} &= (\mu_w^2{+}\sigma_w^2) \cdot v_x + (\mu_x^2{+}\sigma_x^2) \cdot v_w \label{eq:mul-var} \\
c_{xw} &= \mu_w \cdot c_x + \mu_x \cdot c_w \label{eq:mul-cov}
\vspace{-4pt}
\end{align}

\subsubsection{Division}

Division $f(x, w) = x / w$ is decomposed into a single-field reciprocal followed by a cross-field multiplication.

\paragraph{Reciprocal.}
Applying the nonlinear rule (\S\ref{sec:singlefield}) to $g(w) = 1/w$ with $g'(w) = -1/w^2$ and $g''(w) = 2/w^3$ yields the complete output tuple:
\vspace{-4pt}
\begin{equation}
T_{1/w} = \biggl(\frac{1}{\mu_w},\;\;
\frac{\sigma^2_w}{\mu_w^4},\;\;
\frac{-b_w}{\mu_w^2} {+} \frac{v_w}{\mu_w^3} {+} \frac{2c_w}{\mu_w^3},\;\;
\frac{v_w}{\mu_w^4},\;\;
\frac{-c_w}{\mu_w^2}\biggr)
\label{eq:recip-tuple}
\vspace{-4pt}
\end{equation}

Each component follows from Eqs.~\eqref{eq:delta-mu}--\eqref{eq:delta-cov} by substituting the derivatives of $g(w) = 1/w$.

\paragraph{Composition.}
Division follows by multiplying $T_x$ with $T_{1/w}$ using Eqs.~\eqref{eq:mul-bias}--\eqref{eq:mul-cov}, inheriting the same cross-field independence assumption.
The output is again a standard five-tuple, ready for the next operator.

\subsection{Aggregation and Spatial Error Correlation}
\label{sec:aggregation}

The operators above are all pixel-wise: they transform values at each spatial location independently.
Aggregation operators (sum, mean, weighted sum over a spatial block) combine values \emph{across} locations, and the result depends critically on the spatial relationships among errors at different pixels within the same field.

\subsubsection{Why Aggregation Requires Special Treatment}

Consider the sum $S = \sum_{i=1}^{n} e_i$ of per-pixel errors within a block of $n$ pixels.
If the errors were independent, $\mathrm{Var}(S) = n \cdot v$.
In practice, compression errors exhibit strong spatial correlation, and the actual $\mathrm{Var}(S)$ can exceed $n \cdot v$ by orders of magnitude (Table~\ref{tab:alpha-values}).

This correlation arises from two sources.
First, the compressor mechanism introduces shared structure: prediction-based compressors like SZ3 share prediction context across neighboring pixels, while transform-based compressors like SPERR and ZFP share wavelet or DCT coefficients within transform blocks.
In both cases, nearby pixels inherit errors from a common source.
Second, the data structure itself contributes: smooth regions produce similar error pattern, leading to correlated quantization errors even when the compressor does not explicitly couple pixels.

This situation is fundamentally different from the cross-field independence assumed in \S\ref{sec:multifield}.
When two fields are multiplied pixel-wise ($x_i \cdot w_i$), the errors $e_{x,i}$ and $e_{w,i}$ at the same pixel come from independent compression runs and are uncorrelated.
But when a single field is summed over a block ($\sum_i e_i$), the errors $e_i$ and $e_j$ at different pixels come from the \emph{same} compression run and share spatial structure; their element-wise product is position-dependent, not exchangeable.
Treating them as independent would severely underestimate the variance of the aggregated error.

The exact variance involves $O(n^2)$ pairwise covariances:
\begin{equation}
\mathrm{Var}(S) = n \cdot v + n(n{-}1) \cdot \bar{c}
\label{eq:var-identity-simplified}
\end{equation}
where $\bar{c}$ is the mean pairwise covariance.
Computing $\bar{c}$ directly is infeasible for large blocks, so we need a parsimonious model that captures the effect of spatial correlation without tracking $O(n^2)$ pairwise terms.

\subsubsection{The $v_u / v_c$ Decomposition}

Our approach decomposes the per-pixel error variance into two components: an \emph{uncorrelated} part $v_u$ that averages away under summation (like i.i.d.\ noise) and a \emph{correlated} part $v_c$ that persists (like a common-mode offset shared by all pixels in a block).
Writing $v = v_u + v_c$, the block-sum variance becomes:
\begin{equation}
\mathrm{Var}(S) = n \cdot v_u + n^2 \cdot v_c
\label{eq:uv-var}
\end{equation}

Here $v_u$ captures pixel-level noise that averages away under summation (variance scales as $n$), while $v_c \approx \bar{c}$ captures the common-mode signal that does not cancel under summation (variance scales as $n^2$).

\subsubsection{Calibrating $v_u$ and $v_c$ via $\alpha$}
\label{sec:alpha}

We estimate $v_u$ and $v_c$ without computing pairwise covariances.
Partition the error field into non-overlapping blocks of $n$ pixels each, and define the \emph{spatial correlation index}:
\begin{equation}
\alpha = \frac{\mathrm{Var}\!\bigl(\sum_{\text{block}} e_i\bigr)}{n \cdot \mathrm{Var}(e_i)}
\label{eq:alpha-def}
\end{equation}

Under i.i.d.\ errors, $\alpha = 1$; spatial correlation gives $\alpha \gg 1$.
Under the model~\eqref{eq:uv-var}, $\alpha = 1 + (n{-}1) \cdot v_c / v$, yielding:
\vspace{-4pt}
\begin{equation}
v_c = \frac{(\alpha - 1)\,v}{n - 1}, \qquad v_u = v - v_c
\label{eq:alpha-solve}
\vspace{-4pt}
\end{equation}

Crucially, the ratio $r = v_c / v = (\alpha - 1)/(n - 1)$ is approximately \emph{scale-invariant}: once estimated at a reference block size $n_0$, it transfers to other block sizes via $\alpha_{\text{new}} = 1 + (n_{\text{new}} {-} 1) \cdot r$, without re-measurement.
TOPIQ computes $\alpha$ once during offline metadata extraction and reuses it for arbitrary query block sizes at runtime; QoI-preserving compressors, by contrast, require the analysis region to be specified at compression time.

\subsubsection{Propagation of $v_u / v_c$ Through Pixel-wise Operators}

The $v_u / v_c$ decomposition propagates through pixel-wise operators by the same scaling rules derived in \S\ref{sec:singlefield}--\ref{sec:multifield}.
Since pixel-wise operators apply the same transformation at every pixel, they scale both components equally and neither create nor destroy spatial correlations:
\begin{align}
v_u^{(f)} &= \bigl[f'(\mu)\bigr]^2 \cdot v_u, \\
v_c^{(f)} &= \bigl[f'(\mu)\bigr]^2 \cdot v_c.
\end{align}

For cross-field multiplication (Eq.~\ref{eq:mul-var}):
\begin{align}
v_u^{(xw)} &= (\mu_w^2{+}\sigma_w^2) \cdot v_u^{(x)} + (\mu_x^2{+}\sigma_x^2) \cdot v_u^{(w)} \\
v_c^{(xw)} &= (\mu_w^2{+}\sigma_w^2) \cdot v_c^{(x)} + (\mu_x^2{+}\sigma_x^2) \cdot v_c^{(w)}
\end{align}

\subsubsection{Aggregation With $v_u / v_c$}

After propagating $v_u$ and $v_c$ through all pixel-wise operations, the aggregation step uses both components.
For the block sum $S = \sum_{i=1}^n \delta_i$, where $\delta_i$ is the per-pixel output error, the output tuple is:
\vspace{-4pt}
\begin{equation}
T_{\sum} = \bigl(n\mu,\;\; n\sigma^2,\;\; nb,\;\; n \cdot v_u + n^2 \cdot v_c,\;\; nc\bigr)
\label{eq:uv-agg}
\vspace{-4pt}
\end{equation}

The key entry is the error variance: $v_u$ scales linearly (as for independent errors) while $v_c$ scales quadratically (as for perfectly correlated errors), capturing the full range of spatial correlation.
For the block mean, each component is divided by $n$ (variance by $n^2$), giving $v_{\bar{\delta}} = v_u / n + v_c$.

For a weighted sum $\sum_i w_i \delta_i$ with known weights, writing $s_1 = \sum_i w_i$ and $s_2 = \sum_i w_i^2$:
\vspace{-4pt}
\begin{equation}
T_{\sum w} = \bigl(s_1 \mu,\;\; s_2 \sigma^2,\;\; s_1 b,\;\; s_2\, v_u + s_1^2\, v_c,\;\; s_1 c\bigr)
\label{eq:weighted-sum-var}
\vspace{-4pt}
\end{equation}

The $v_u$ term is weighted by $s_2 = \sum w_i^2$ (each pixel contributes independently), while $v_c$ is weighted by $s_1^2 = (\sum w_i)^2$ (the common-mode contribution scales with the total weight).
The output can feed into further operators; for the block sizes used in our evaluation ($200{\times}200$ for 2D, $32^3$ for 3D), this model is well-calibrated (\S\ref{sec:accuracy}).

\subsection{Composing QoIs from Primitive Operators}
\label{sec:composing-qoi}
Any QoI expressible as a directed acyclic graph of the above \emph{smooth} operators can be handled by composing the corresponding propagation rules.
At each node the input tuple(s) are transformed into an output tuple; the final node yields the QoI-level bias and variance.
No per-QoI derivation or retraining.
The retained first- and second-order terms are sufficient to propagate bias and variance accurately across multi-step pipelines: in our evaluation, longer multi-field QoIs (e.g., the six-operator cloudy ratio) achieve calibration comparable to single-operator QoIs.
Non-smooth operators such as ReLU, max, min, and threshold-based feature counts fall outside this scope; they lack useful second derivatives and would require fundamentally different propagation techniques (see \S\ref{sec:conclusion}).

Figure~\ref{fig:qoi-graphs} illustrates two examples evaluated in \S\ref{sec:accuracy}:
a single-field neural network QoI that chains weighted-sum aggregation, sigmoid, and linear operators,
and a multi-field cloudy ratio that combines subtraction, sigmoid masking, multiplication, summation, and division across three compressed fields.
Both go beyond what existing QoI-preserving compressors support: QPET~\cite{liu2025vldb} targets univariate symbolic functions and simple multivariate expressions but does not natively handle multi-layer neural networks or multi-field ratio QoIs.
TOPIQ composes these from the same primitive operators without case-specific derivation.

\begin{figure}[t]
\centering
\small
\begin{tcolorbox}[colback=white, colframe=black, boxrule=0.4pt, arc=0mm,
  top=1mm, bottom=1mm, left=2mm, right=2mm,
  title={\footnotesize\textbf{(a) Block-input neural network:} $q = W_2\,\sigma(W_1 \mathbf{x} + b_1) + b_2$,\; $\mathbf{x} \in \mathbb{R}^n$}]
\footnotesize
\begin{tabular}{@{}r@{$\;\to\;$}l@{\quad}l@{}}
$T_x$ & input error tuple & $(\mu, \sigma^2, b, v_u, v_c, c)$ \\
$W_1 \mathbf{x} {+} b_1$ & \texttt{weighted\_sum} (\S\ref{sec:aggregation}) & Eq.~\eqref{eq:weighted-sum-var} \\
$\sigma(\cdot)$ & \texttt{sigmoid} (\S\ref{sec:singlefield}) & Eqs.~\eqref{eq:delta-bias}--\eqref{eq:delta-var} \\
$W_2 (\cdot) {+} b_2$ & \texttt{linear} (\S\ref{sec:singlefield}) & Eq.~\eqref{eq:linear-tuple} \\
\end{tabular}
\end{tcolorbox}
\vspace{2mm}
\begin{tcolorbox}[colback=white, colframe=black, boxrule=0.4pt, arc=0mm,
  top=1mm, bottom=1mm, left=2mm, right=2mm,
  title={\footnotesize\textbf{(b) Cloudy ratio:} $q = \frac{\sum m_i (F_i^c {-} F_i)}{\sum m_i + \epsilon}$,\; $m_i = \sigma(k(\text{CLDTOT}_i {-} \tau))$}]
\footnotesize
\begin{tabular}{@{}r@{$\;\to\;$}l@{\quad}l@{}}
$T_{\text{FLUTC}} - T_{\text{FLUT}}$ & \texttt{sub} (\S\ref{sec:multifield}) & Eq.~\eqref{eq:add-two-field} \\
$k\,(T_{\text{CLDTOT}} - \tau)$ & \texttt{linear} (\S\ref{sec:singlefield}) & Eq.~\eqref{eq:linear-tuple} \\
$\sigma(\cdot)$ & \texttt{sigmoid} (\S\ref{sec:singlefield}) & Eqs.~\eqref{eq:delta-bias}--\eqref{eq:delta-var} \\
$T_m \cdot T_X$ & \texttt{multiply} (\S\ref{sec:multiplication}) & Eqs.~\eqref{eq:mul-bias}--\eqref{eq:mul-cov} \\
$\sum / \sum$ & \texttt{aggregate} $\times 2$ (\S\ref{sec:aggregation}) & Eq.~\eqref{eq:uv-agg} \\
$\mathrm{num} / (\mathrm{den} {+} \epsilon)$ & \texttt{div} (\S\ref{sec:multifield}) & Eq.~\eqref{eq:recip-tuple} + multiply \\
\end{tabular}
\end{tcolorbox}
\caption{Two QoIs from \S\ref{sec:accuracy} expressed as operator chains.
Each step transforms input tuple(s) into an output tuple using the rules derived above.
No QoI-specific derivation is needed; the framework composes automatically.}
\label{fig:qoi-graphs}
\vspace{-1em}
\end{figure}
\section{Evaluation}
\label{sec:eval}

\subsection{Setup and Metrics}
\label{sec:setup}

We evaluate on selected fields from four scientific datasets:
CLDTOT, CLDHGH, FLUT, FLUTC from \textbf{CESM-ATM} (2D, $1800 {\times} 3600$, float32);
dark-matter density and baryon density, and $velocity_x$ from \textbf{NYX}~\cite{nyx,sdrbench}(3D, $512^3$, float32);
PRES and T from \textbf{SCALE-LETKF}~\cite{SCALE-LETKF,sdrbench} (3D, $98 {\times} 1200 {\times} 1200$, float32);
TC and U from \textbf{Hurricane-ISABEL}~\cite{sdrbench} (3D, $100 {\times} 500 {\times} 500$, float32).
Three compressors,SZ3~\cite{sz3}, SPERR~\cite{SPERR}, and ZFP~\cite{zfp},are tested in absolute-error-bound mode with $\varepsilon_{\mathrm{rel}} \in \{5 {\times} 10^{-3}, \ldots, 10^{-5}\}$.

Given a predicted QoI $\hat{q}$ with predicted standard deviation $\hat{\sigma}$ and true value $q^*$, the \emph{standardized residual} (z-score) is $z = (q^* - \hat{q}) / \hat{\sigma}$.
Well-calibrated predictions yield $z \sim \mathcal{N}(0,1)$; we report the bias $\mu_z$ (ideal~0), the calibration metric $\sigma_z$ (ideal~1), and coverage at $3\sigma$ (ideal $99.7\%$).
All these evaluations run on a compute node from an HPC cluster. The node is configured with dual-socket Intel Xeon E5-2695 v4 processors, providing a total of 36 physical cores (18 per processor) per node, and 128GB DDR4 memory.

\subsection{Ablation Study}
\label{sec:ablation}
Before evaluating end-to-end prediction accuracy, we first identify which modeling choices in TOPIQ are critical, as these assumptions directly determine the validity of the resulting uncertainty estimates.
TOPIQ's propagation rules make several independence assumptions for tractability, most notably cross-field independence (\S\ref{sec:multiplication}).
However, two specific dependencies \emph{cannot} be dropped without catastrophic degradation: (1)~spatial correlation between error pixels, captured by the correlation index~$\alpha$, and (2)~within-field data--error coupling, captured by $\mathrm{Cov}(x,e)$.
We validate this with three ablation modes: ignore $\alpha$ ($\alpha{=}1$), ignore $\mathrm{Cov}(x,e)$ ($c{=}0$), and a decorrelated control (random pixel permutation).

\subsubsection{Spatial Correlation Is Pervasive and Consequential}

We first examine whether compression errors are spatially correlated and, if so, whether ignoring this correlation degrades predictions.
The correlation index $\alpha$ varies by three orders of magnitude across compressors and datasets (Table~\ref{tab:alpha-values}), ranging from $\alpha \approx 1$ (i.i.d.) to $\alpha > 700$.
Randomly permuting error pixels collapses $\alpha$ to ${\approx}1$ in every case, confirming the correlation is spatial.
Ignoring this correlation has severe consequences: on ZFP/CLDTOT ($\alpha = 218$), setting $\alpha{=}1$ inflates the calibration metric~$\sigma_z$ from $1.0$ to $14.7$, underestimating aggregation variance by an order of magnitude.
When $\alpha$ is small (e.g., SPERR/CLDTOT, $\alpha \approx 1.7$), ignoring correlation has minimal effect, confirming that the model correctly reduces to the i.i.d.\ case.

\begin{table}[t]
\centering
\caption{The spatial correlation index $\alpha$ varies by three orders of magnitude across compressors, datasets, and fields.
$\alpha{=}1$ corresponds to i.i.d.\ errors; $\alpha \gg 1$ indicates spatial correlation that inflates block-aggregated variance.
$\alpha_{\mathrm{shuf}} \approx 1$ after random permutation confirms the effect is spatial.}
\label{tab:alpha-values}
\small
\begin{tabular}{llrr}
\toprule
Dataset & Compressor / Field & $\alpha$ & $\alpha_{\mathrm{shuf}}$ \\
\midrule
\multirow{6}{*}{\rotatebox{90}{CESM}}
& SZ3 / CLDTOT   &     2.0 & 1.02 \\
& SZ3 / CLDHGH   &     3.6 & 1.00 \\
& SPERR / CLDTOT &     1.7 & 0.87 \\
& SPERR / CLDHGH &     1.7 & 1.05 \\
& ZFP / CLDTOT   & \textbf{217.5} & 0.86 \\
& ZFP / CLDHGH   & \textbf{119.9} & 0.85 \\
\midrule
\multirow{3}{*}{\rotatebox{90}{NYX}}
& SZ3 / velocity\_x  &  123.0 & 1.01 \\
& SPERR / velocity\_x &  10.0 & 0.99 \\
& ZFP / velocity\_x   &  46.2 & 1.00 \\
\midrule
\multirow{6}{*}{\rotatebox{90}{SCALE}}
& SZ3 / PRES   & \textbf{4111.3} & 1.05 \\
& SZ3 / T      & \textbf{673.7} & 0.98 \\
& SPERR / PRES &  79.5 & 0.99 \\
& SPERR / T    &  16.4 & 1.00 \\
& ZFP / PRES   & \textbf{179.2} & 1.00 \\
& ZFP / T      & 144.2 & 1.01 \\
\midrule
\multirow{5}{*}{\rotatebox{90}{Hurr.}}
& SZ3 / TC     &  58.9 & 1.05 \\
& SZ3 / U      &  12.2 & 1.01 \\
& SPERR / TC   &   7.1 & 0.93 \\
& ZFP / TC     &  17.1 & 0.99 \\
& ZFP / U      &  23.4 & 0.93 \\
\bottomrule
\end{tabular}
\vspace{-1em}
\end{table}

\subsubsection{Data--Error Coupling Introduces Systematic Bias}

\begin{figure}[t]
\centering
\includegraphics[width=\linewidth]{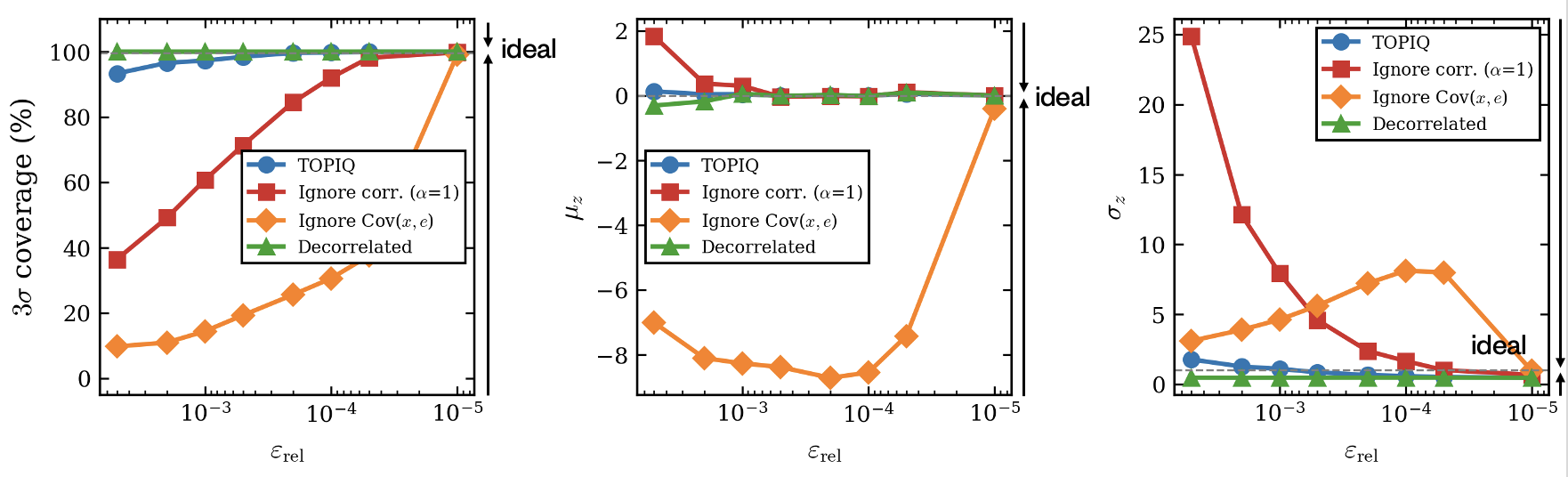}
\caption{TOPIQ correctly predicts QoI uncertainty where na\"ive independence assumptions fail.
\textbf{Ablation on mean$(x^2)$} (NYX dark-matter density, SPERR):
ignoring the correlation index~$\alpha$ (red) inflates the calibration metric~$\sigma_z$ to $7.9$;
ignoring the data--error covariance (orange) shifts the bias~$\mu_z$ to $-8.72$.
TOPIQ (blue) corrects both effects.
Randomly permuting errors (green) restores $\sigma_z \approx 1$, confirming both effects are spatial.
Both ablation lines recover at tight bounds as SPERR's spatial correlation diminishes.}
\vspace{-1em}
\label{fig:ablation}
\end{figure}

Beyond spatial correlation, we analyze the impact of the coupling between data values and their compression errors, as this interaction directly affects prediction bias.
Ignoring the data--error covariance $\mathrm{Cov}(x,e)$ shifts the bias metric~$\mu_z$ rather than inflating $\sigma_z$.
Figure~\ref{fig:ablation} demonstrates both effects on dark matter density field in NYX dataset with SPERR (correlation index $\alpha \approx 53$):
ignoring $\alpha$ (red) inflates $\sigma_z$ to $7.9$; ignoring $\mathrm{Cov}(x,e)$ (orange) shifts $\mu_z$ to $-8.72$.
TOPIQ (blue) corrects both; the decorrelated control (green) restores $\sigma_z \approx 1$ by eliminating spatial structure.

Both ablation lines recover at very tight bounds ($\varepsilon_{\mathrm{rel}} \lesssim 10^{-4}$): SPERR's correlation index drops from $\alpha = 53$ to $2$ and $|\mathrm{Cov}(x,e)|$ diminishes by four orders of magnitude.
This trend is compressor-specific.

\paragraph{Summary.}
TOPIQ's independence assumptions hold in most cases.
The ablation identifies two that do not: (1)~spatial error correlation inflates $\sigma_z$ by up to $14.7{\times}$; (2)~data--error coupling shifts $\mu_z$ by up to $-8.72$.
Both are purely spatial: permuting error pixels eliminates both effects.

\subsection{Prediction Accuracy}
\label{sec:accuracy}
Having established that spatial correlation and data-error coupling are essential, we now evaluate whether TOPIQ's full model produces well-calibrated uncertainty predictions across diverse datasets, compressors, and QoI types.
TOPIQ achieves well-calibrated uncertainty estimates across 4 datasets, 3 compressors, and 4 QoI families without per-QoI tuning.
We evaluate four QoI families of increasing complexity, each expressed as a composition of TOPIQ's primitive operators.
Notably, QoI families (2)--(4) go beyond what most existing QoI-preserving compressors readily support: frameworks such as QPET~\cite{liu2025vldb} primarily target univariate symbolic functions (e.g., $x^2$, $\sin(x)$) with limited multivariate support (e.g., vector magnitude), but do not natively handle multi-layer neural networks, general cross-field products, or multi-field ratio QoIs.
TOPIQ supports all of these through operator composition:

\textbf{(1)~Mean Square} (single field, aggregation + nonlinear):
\vspace{-4pt}
\begin{equation}
  q_1 = \frac{1}{n}\sum_{i=1}^n x_i^2.
  \label{eq:qoi-ms}
\vspace{-4pt}
\end{equation}
Operator chain: $\texttt{square} \to \texttt{mean}$.

\textbf{(2)~Block-input neural network} (single field, 1-hidden-layer MLP, 1.28M parameters):
We have introduced it in \S\ref{sec:composing-qoi}.
The first layer is an aggregation (weighted sum over pixels with learned weights), making this QoI fundamentally different from pointwise functions: it cannot be expressed as a per-pixel symbolic expression.
We include this case as a stress test: it represents a complex post-hoc QoI that existing QoI-preserving compressors are not designed to preserve directly.
TOPIQ currently supports smooth operator chains, so networks with non-smooth activations (e.g., ReLU) are outside the present scope; extending to such cases via piecewise or distribution-shape propagation is left to future work.

\textbf{(3)~Weighted sum} (two compressed fields $x$ and $w$):
\vspace{-4pt}
\begin{equation}
  q_3 = \sum_{i=1}^n x_i \cdot w_i.
  \label{eq:qoi-ws}
\vspace{-4pt}
\end{equation}
Operator chain: $\texttt{multiply}_{x,w} \to \texttt{sum}$.
This exercises the cross-field multiplication rule (\S\ref{sec:multiplication}).

\textbf{(4)~Cloudy ratio} (three fields: FLUT, FLUTC, CLDTOT; CESM only):
This quantity is a physically motivated proxy for regional cloud--radiative analysis. In climate analysis, cloud radiative effects and their spatially aggregated contributions are standard diagnostics for understanding how clouds modulate top-of-atmosphere radiation~\cite{ramanathan1989cre,zelinka2012kernels}. Our formulation is a simplified and differentiable version of this idea: instead of using a hard cloudy/clear partition, we apply a sigmoid gate as a smooth approximation to a binary cloud-fraction threshold. The implementation can be found in \S\ref{sec:composing-qoi}. The hyper parameters are, $k{=}20$, $\tau{=}0.5$, and $\epsilon{=}10^{-6}$.

Operator chain: $\texttt{sub} \to \texttt{mul} \to \texttt{sigmoid} \to \texttt{mul} \to \texttt{sum} \to \texttt{div}$.

Table~\ref{tab:prediction-accuracy} summarizes 69 configurations at $\varepsilon_{\mathrm{rel}}{=}10^{-3}$: 65 (94.2\%) achieve $\sigma_z \in [0.7, 1.3]$.
Coverage matches targets (median observed: $99.6\%$).
Some compressors introduce systematic bias; TOPIQ propagates the measured global bias unconditionally (\S\ref{sec:tuple}).

\begin{table*}[t]
\centering
\caption{\textbf{Prediction accuracy at $\varepsilon_{\mathrm{rel}}{=}10^{-3}$.}
A representative subset of 69 configurations, selected to cover every dataset, compressor, and QoI family at least once while including both strong and weak cases.
The 40 omitted configurations follow the same pattern; across all 69, 65 (94.2\%) achieve $\sigma_z \in [0.7, 1.3]$.
Cell shading:
\colorbox{dg}{\strut excellent} (${\leq}0.05$)\,/\,\colorbox{lg}{\strut good} (${\leq}0.10$)\,/\,\colorbox{ly}{\strut marginal} (${\leq}0.20$)\,/\,\colorbox{lr}{\strut poor} (${\leq}0.50$)\,/\,\colorbox{dr}{\strut failure} (${>}0.50$)
for $|\mu_z|$; same structure with thresholds $0.05$/$0.15$/$0.30$/$0.50$ for $|\sigma_z{-}1|$; and $3$/$5$/$10$/$15$ percentage points for coverage.
$\dagger$: under-calibrated cases discussed in \S\ref{sec:limitations}.
$\ddagger$: over-conservative ($\sigma_z < 1$), discussed in \S\ref{sec:limitations}.}
\label{tab:prediction-accuracy}
\small
\begin{tabular}{ll l rrr}
\toprule
Dataset & Compressor & QoI / Field & $\mu_z$ & $\sigma_z$ & cov$_{3\sigma}$ \\
\midrule
\multirow{6}{*}{CESM}
& SZ3    & mean$(x^2)$ / CLDTOT                   & \cellcolor{dg} +0.00 & \cellcolor{dg} 1.01 & \cellcolor{dg} 98.8 \\
& SPERR  & neural net / CLDTOT                    & \cellcolor{lg} +0.08 & \cellcolor{dg} 0.97 & \cellcolor{dg} 100.0 \\
& ZFP    & mean$(x^2)$ / CLDTOT                   & \cellcolor{dg} +0.01 & \cellcolor{dg} 1.00 & \cellcolor{dg} 100.0 \\
& SZ3    & weighted sum / CLDTOT, CLDHGH           & \cellcolor{lg} +0.06 & \cellcolor{dg} 1.04 & \cellcolor{dg} 98.1 \\
& SZ3    & cloudy ratio / FLUT, FLUTC, CLDTOT       & \cellcolor{lg} +0.05 & \cellcolor{lg} 0.92 & \cellcolor{dg} 100.0 \\
& ZFP    & cloudy ratio$^\ddagger$ / FLUT, FLUTC, CLDTOT & \cellcolor{dg} $-$0.00 & \cellcolor{dg} 0.48 & \cellcolor{dg} 100.0 \\
\midrule
\multirow{6}{*}{Hurricane}
& SZ3    & mean$(x^2)$ / TC                       & \cellcolor{lr} $-$0.25 & \cellcolor{lg} 0.91 & \cellcolor{dg} 98.7 \\
& SPERR  & mean$(x^2)$ / TC                       & \cellcolor{dg} +0.02 & \cellcolor{dg} 1.01 & \cellcolor{dg} 98.7 \\
& ZFP    & neural net / U                         & \cellcolor{dg} $-$0.03 & \cellcolor{dg} 1.01 & \cellcolor{dg} 99.9 \\
& SZ3    & neural net / TC                        & \cellcolor{lg} +0.06 & \cellcolor{dg} 1.00 & \cellcolor{dg} 99.7 \\
& SZ3    & weighted sum / TC, U                    & \cellcolor{ly} +0.15 & \cellcolor{lg} 1.13 & \cellcolor{lg} 97.8 \\
& ZFP    & weighted sum / TC, U                    & \cellcolor{dg} +0.01 & \cellcolor{lg} 0.93 & \cellcolor{dg} 100.0 \\
\midrule
\multirow{7}{*}{NYX}
& SZ3    & mean$(x^2)$ / velocity\_x               & \cellcolor{dg} $-$0.04 & \cellcolor{dg} 1.00 & \cellcolor{dg} 99.2 \\
& SPERR  & mean$(x^2)$ / velocity\_x               & \cellcolor{dg} +0.01 & \cellcolor{dg} 1.02 & \cellcolor{dg} 98.8 \\
& ZFP    & neural net / velocity\_x                & \cellcolor{dg} $-$0.02 & \cellcolor{dg} 1.00 & \cellcolor{dg} 99.8 \\
& SZ3    & neural net / dark\_matter\_density         & \cellcolor{dg} +0.00 & \cellcolor{dg} 1.00 & \cellcolor{dg} 99.7 \\
& SPERR  & neural net / dark\_matter\_density         & \cellcolor{dg} $-$0.01 & \cellcolor{dg} 1.00 & \cellcolor{dg} 99.8 \\
& ZFP    & neural net / baryon\_density              & \cellcolor{dg} $-$0.02 & \cellcolor{dg} 1.01 & \cellcolor{dg} 99.7 \\
& SZ3    & weighted sum / dark\_matter\_density, velocity\_x & \cellcolor{dg} $-$0.00 & \cellcolor{lg} 0.89 & \cellcolor{dg} 99.7 \\
\midrule
\multirow{10}{*}{SCALE}
& SZ3    & mean$(x^2)$ / PRES                     & \cellcolor{dg} $-$0.01 & \cellcolor{lg} 1.09 & \cellcolor{dg} 100.0 \\
& SZ3    & mean$(x^2)$ / T                        & \cellcolor{dg} $-$0.05 & \cellcolor{lg} 1.10 & \cellcolor{dg} 97.2 \\
& SZ3    & neural net / PRES                      & \cellcolor{lg} $-$0.08 & \cellcolor{lg} 0.91 & \cellcolor{dg} 99.8 \\
& ZFP    & mean$(x^2)$ / PRES                     & \cellcolor{dg} +0.01 & \cellcolor{dg} 1.01 & \cellcolor{dg} 100.0 \\
& ZFP    & mean$(x^2)$ / T                        & \cellcolor{dg} +0.00 & \cellcolor{dg} 1.04 & \cellcolor{dg} 100.0 \\
& SPERR  & mean$(x^2)$ / PRES$^\dagger$           & \cellcolor{dg} +0.05 & \cellcolor{dr} 1.76 & \cellcolor{ly} 92.5 \\
& SPERR  & mean$(x^2)$ / T$^\dagger$              & \cellcolor{dg} +0.01 & \cellcolor{lr} 1.36 & \cellcolor{lg} 96.7 \\
& SPERR  & neural net / PRES                      & \cellcolor{lg} +0.07 & \cellcolor{dg} 1.03 & \cellcolor{dg} 99.6 \\
& SZ3    & weighted sum / PRES, T                  & \cellcolor{dg} +0.03 & \cellcolor{ly} 1.19 & \cellcolor{dg} 99.9 \\
& SPERR  & weighted sum$^\dagger$ / PRES, T       & \cellcolor{lg} +0.06 & \cellcolor{dr} 1.89 & \cellcolor{ly} 91.5 \\
\bottomrule
\end{tabular}
\vspace{-1em}
\end{table*}

\subsubsection{Calibration Across Error Bounds}

\begin{figure}[t]
\centering
\includegraphics[width=\linewidth]{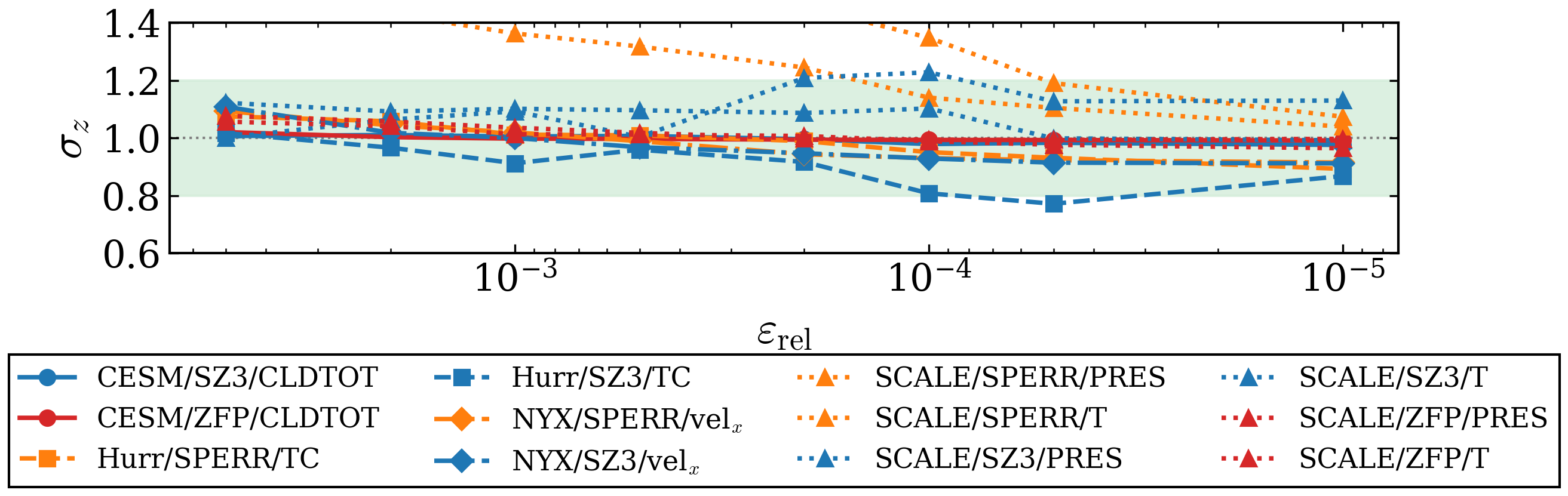}
\caption{TOPIQ maintains calibration across two orders of magnitude in error bound.
Shown for the mean$(x^2)$ QoI on twelve dataset/compressor combinations; other QoI families exhibit similar trends.
The calibration metric~$\sigma_z$ stays within the specified confidence interval $[0.8, 1.2]$ (shaded band) from $\varepsilon_{\mathrm{rel}}{=}5{\times}10^{-3}$ to $10^{-5}$.}
\vspace{-1em}
\label{fig:sigma-z-sweep}
\end{figure}

Calibration is stable across two orders of magnitude in error bound.
Figure~\ref{fig:sigma-z-sweep} shows the calibration metric~$\sigma_z$ for twelve representative dataset/compressor combinations on the mean$(x^2)$ QoI: all remain within $[0.8, 1.2]$ from $\varepsilon_{\mathrm{rel}}{=}5{\times}10^{-3}$ to $10^{-5}$; other QoI families exhibit similar stability.
Across all 552 evaluations (69 configs $\times$ 8 error bounds), 514 (93.1\%) fall within $\sigma_z \in [0.7, 1.3]$, with failures concentrated in SCALE-LETKF (\S\ref{sec:limitations}).

\subsubsection{Limitations}
\label{sec:limitations}

At $\varepsilon_{\mathrm{rel}}{=}10^{-3}$, 3 of 69 configurations are under-calibrated ($\sigma_z > 1.3$), attributable to spatial non-stationarity; one additional configuration is over-conservative ($\sigma_z < 1$), which we discuss separately.

\begin{figure}[]
\centering
\begin{subfigure}[]{0.3\linewidth}
  \centering
  \includegraphics[width=\linewidth]{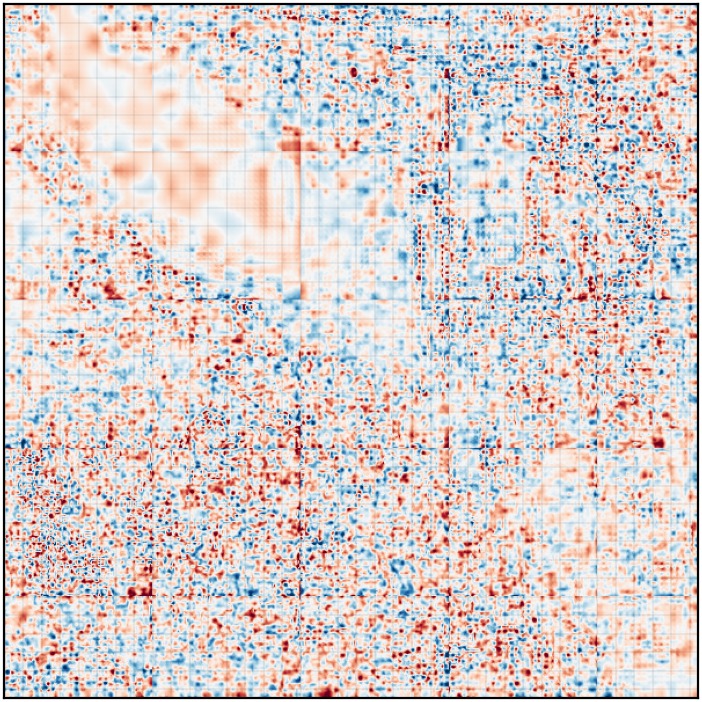}
  \caption{PRES / SPERR}
\end{subfigure}
\hspace{2em}
\begin{subfigure}[]{0.3\linewidth}
  \centering
  \includegraphics[width=\linewidth]{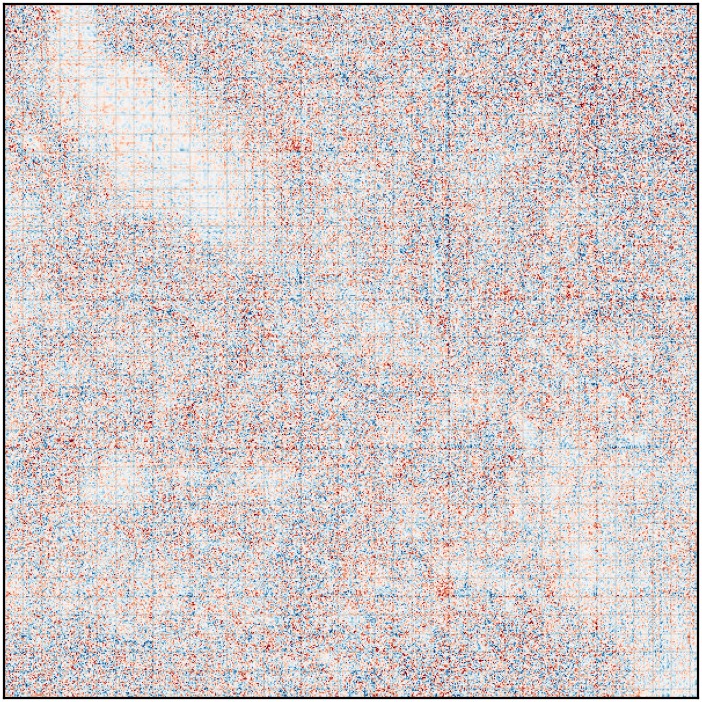}
  \caption{T / SPERR}
\end{subfigure}
\caption{Compression error fields on SCALE-LETKF with SPERR (slice $z{=}95$, $\varepsilon_{\mathrm{rel}}{=}10^{-3}$, $32{\times}32$ block grid; each subplot uses its own color scale).
(a)~PRES: SPERR's wavelet decomposition produces spatially non-stationary errors; smooth blocks have tiny errors while high-gradient blocks produce large structured artifacts (block-internal std varies $939{\times}$), causing $\sigma_z \approx 1.8$.
(b)~T: the same compressor produces uniform errors, and TOPIQ achieves $\sigma_z \approx 1.0$.
Guidance: when per-block error variance is highly heterogeneous (e.g. with SPERR), consider switching to a compressor with more spatially uniform error distribution or using per-region $\alpha$.}
\label{fig:error-vis}
\vspace{-1em}
\end{figure}

\textbf{(1)~Spatial non-stationarity of compression errors} (3 configs, all SPERR on SCALE-LETKF).
TOPIQ uses a single global correlation index $\alpha$, which works well when compression errors are spatially stationary.
SPERR's wavelet decomposition produces errors whose spatial structure depends strongly on the local data complexity: smooth regions yield tiny, nearly uncorrelated errors, while high-gradient regions produce large, structured errors that inherit the wavelet's spatial support.
On SCALE-LETKF PRES, this effect is extreme because the field itself has pronounced spatial heterogeneity (block-internal standard deviation varies $939{\times}$, from ${\sim}0.5$ in smooth regions to ${\sim}511$ near pressure fronts), causing per-block $\alpha$ to vary by up to $9{\times}$ (Figure~\ref{fig:error-vis}).
No single global $\alpha$ can represent both regimes, leading to $\sigma_z \approx 1.8$.
By contrast, the T field on the same grid has more uniform structure, and TOPIQ achieves $\sigma_z \approx 1.0$ with all compressors.

\emph{Actionable guidance.}
A high coefficient of variation of per-block error variance serves as a diagnostic flag for this failure mode.
When flagged, users should either (a)~switch to a compressor with more spatially uniform error distribution.

\textbf{(2)~Over-conservative prediction on multi-field ratio QoI} (1 config).
ZFP's cloudy ratio on CESM-ATM yields $\sigma_z \approx 0.50$: systematic biases in individual fields partially cancel in the ratio, causing TOPIQ's predicted uncertainty to exceed the actual error.
From a preservation standpoint, over-conservative predictions are safe: the reported confidence interval is wider than necessary, so the true QoI value is always contained.
More generally, the cross-field independence assumption (\S\ref{sec:multifield}) drops cross-field error correlations, which can make the independent-error estimate either conservative or optimistic.
In our evaluation this effect appeared in 8 of 552 cases, but its frequency is domain-dependent: feature- or event-oriented QoIs (e.g., in fusion or seismology applications) may encounter such coupling more frequently, while the climate- and cosmology-style QoIs in our evaluation showed it less often.
An optional extension is to store cross-field error-correlation metadata for selected field pairs and add the corresponding covariance terms to the multi-field rules; we omit it by default because this changes the metadata from per-field to pairwise summaries.

The under-calibrated cases across all evaluations are concentrated in SPERR on SCALE-LETKF PRES (mechanism~1), confirming they are systematic and data/compressor-specific rather than random.

\subsubsection{Post-Hoc Spatial Queries}
\begin{figure}[]
\centering
\includegraphics[width=\linewidth]{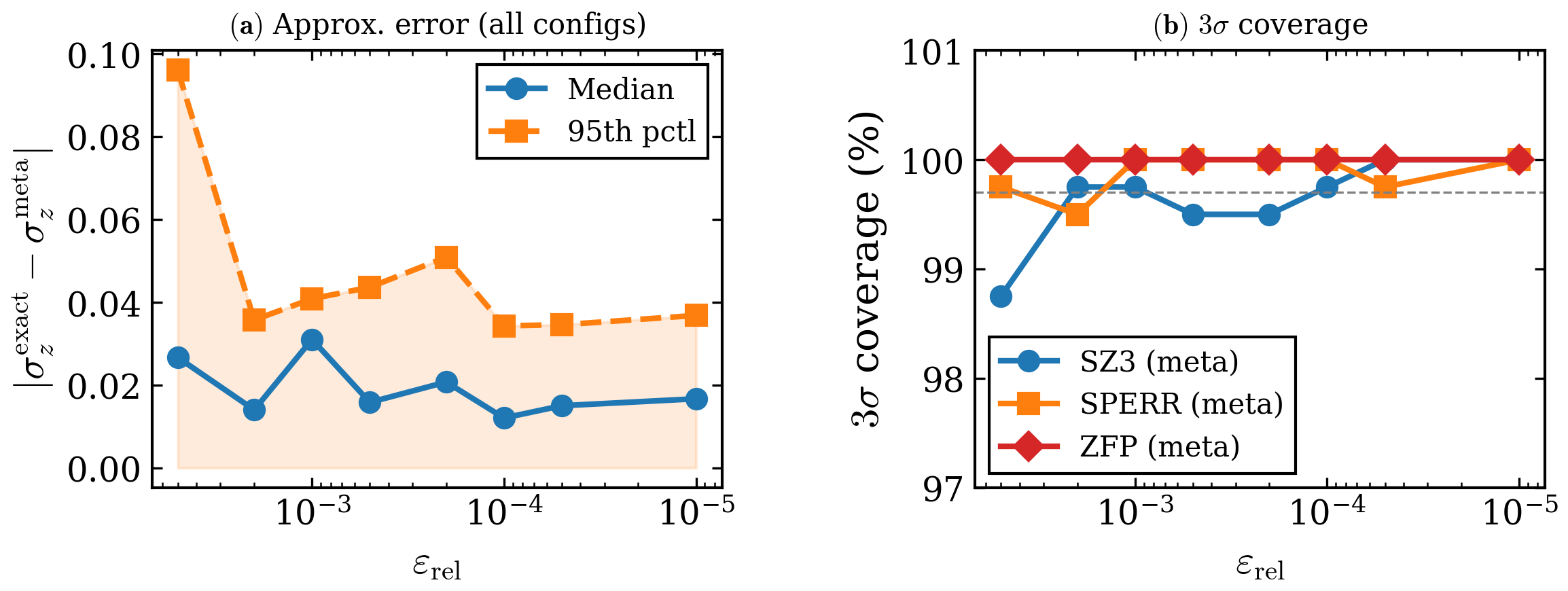}
\caption{Pre-computed metadata enables accurate uncertainty prediction for arbitrary spatial queries on CESM-ATM (200 random $200{\times}200$ regions per configuration).
(a)~Approximation error $|\sigma_z^{\mathrm{exact}} - \sigma_z^{\mathrm{meta}}|$ across all configurations (2 fields $\times$ 3 compressors): median ${<}\,0.03$.
(b)~$3\sigma$ coverage from meta-interpolated predictions almost remains ${\geq}99\%$ across all compressors and error bounds.}
\label{fig:random-block}
\vspace{-1em}
\end{figure}

The evaluations above use block-aligned queries that match the metadata grid exactly. In practice, queries may target arbitrary regions not known at compression time; we now test whether TOPIQ can serve such queries using interpolated metadata.
We pre-compute block-level error metadata on a coarse $150{\times}150$ grid (${\sim}9$\,KB per field on CESM, always $<0.1\%$ of original data) and interpolate to arbitrary query regions via area-weighting.
Finer query granularity requires a denser metadata grid; even at an extremely small query size of $16{\times}16$, the metadata remains ${\approx}1.6\%$ of the field.
Figure~\ref{fig:random-block} evaluates this on CESM-ATM with 200 random $200{\times}200$ regions per configuration:
the meta-interpolated calibration metric~$\sigma_z$ tracks the exact block-local prediction closely (median $|\Delta\sigma_z| < 0.04$), and $3\sigma$ coverage remains ${\geq}99\%$ across all compressors and error bounds.
This capability is essential for workflows where the query region is not known at compression time, such as agent-driven analysis pipelines (\S\ref{sec:casestudy}).

\subsubsection{Prediction Overhead}

\begin{figure}[t]
\centering
\includegraphics[width=0.75\linewidth]{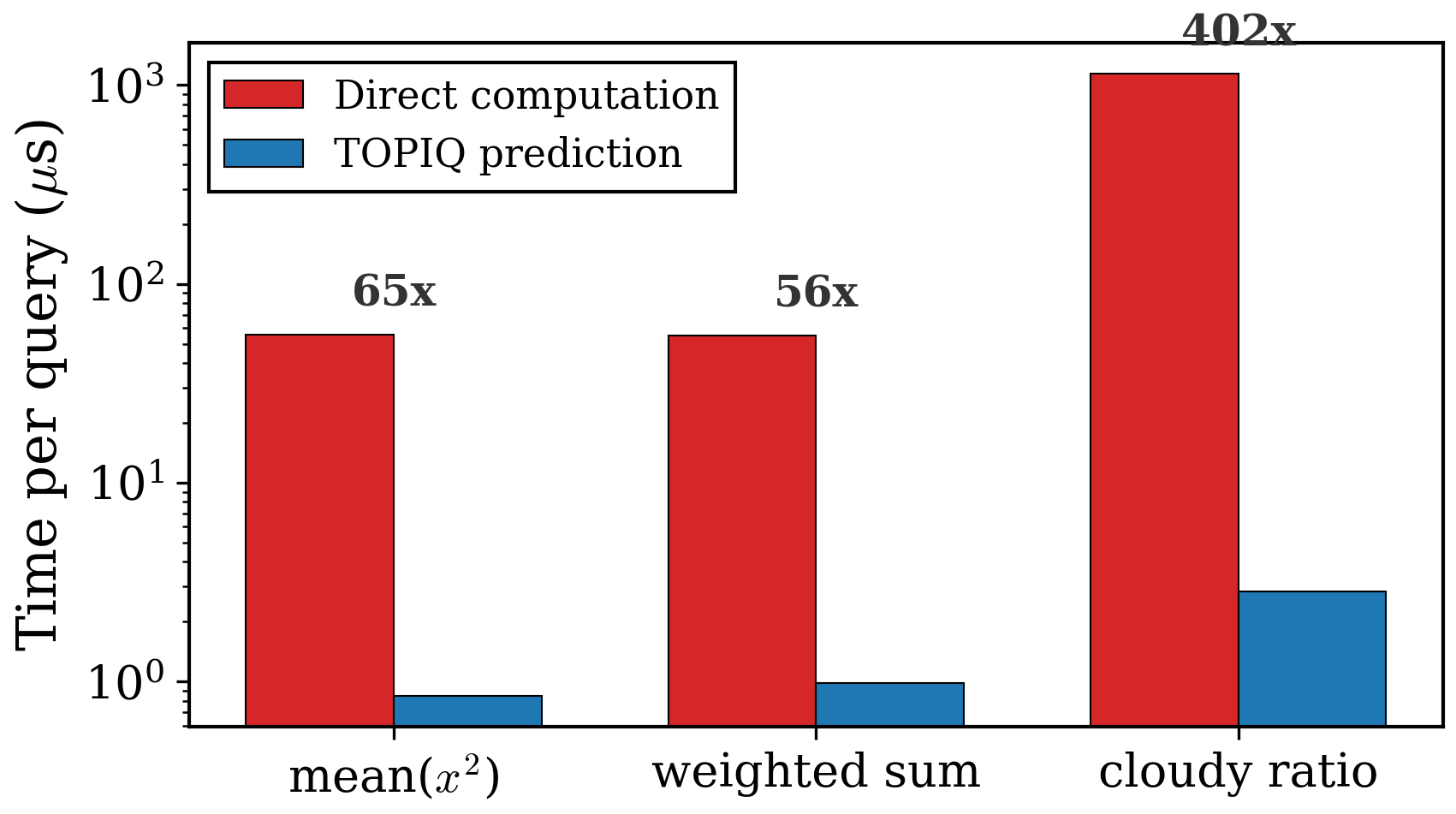}
\caption{TOPIQ prediction (\S\ref{sec:composing-qoi}) is $56{\times}$--$402{\times}$ faster than direct QoI error computation ($200{\times}200$ blocks on CESM-ATM, SZ3, $\varepsilon_{\mathrm{rel}}{=}10^{-3}$).
Once the original data is discarded after compression, direct error computation is no longer possible; TOPIQ requires only pre-computed metadata ($<0.1\%$ of original data) to predict QoI-level uncertainty.}
\label{fig:perf}
\vspace{-1em}
\end{figure}

Accuracy is useful only if predictions are fast enough to be practical.
TOPIQ's prediction requires only $O(1)$ floating-point operations per query (independent of block size), compared to $O(n)$ for direct computation.
Figure~\ref{fig:perf} compares per-query prediction time against direct QoI error computation on $200{\times}200$ blocks.
TOPIQ achieves $56{\times}$--$402{\times}$ speedup on CESM-ATM depending on QoI complexity.
More importantly, direct computation requires both the original and decompressed arrays in memory, whereas TOPIQ requires only pre-computed metadata ($<0.1\%$ of original data, ${\sim}33$\,ms one-time pre-computation per field).
Once the original data is discarded, which is the very purpose of compression, direct error computation becomes impossible; TOPIQ's metadata makes uncertainty quantification feasible after compression.

\section{Case Study: LLM-Driven Climate Analysis}
\label{sec:casestudy}

This case study illustrates how TOPIQ changes the way uncertainty is handled in automated scientific pipelines: rather than requiring QoI definitions at compression time, TOPIQ enables uncertainty quantification for QoIs that are constructed dynamically at query time, without re-compression or QoI-specific derivation.

LLMs are increasingly applied to scientific data analysis, including climate question answering~\cite{jiang2025geogridbench}.
A natural next step is to decompose complex scientific questions into sequences of simple data operations and execute them automatically, which is an approach explored in recent community prototypes~\cite{liu2025beyond}.
Each operation in such a pipeline (read a pixel, compute a regional mean, take a difference) maps directly to a TOPIQ primitive operator; once decomposed, the error propagation follows by composition with no per-query derivation.

\paragraph{Example.}
Consider the question \emph{``How has average 2m temperature changed between January and July in Atlanta?''} (Figure~\ref{fig:scidag-dag}).
An LLM decomposes this into: resolve the city to grid coordinates (metadata, no error), read two monthly pixel values from lossy-compressed storage (data-reading: creates error tuples), and compute their difference (subtraction: doubles the error variance).
The final answer carries a calibrated confidence interval(e.g. $20.9 \pm 3\sqrt{v_\Delta}$\textdegree C at 99.7\%) derived entirely from pre-computed compression metadata, with no re-decompression.

\begin{figure}[t]
\centering
\small
\begin{tcolorbox}[colback=gray!5, colframe=black, boxrule=0.4pt, arc=1mm,
  top=1.5mm, bottom=1.5mm, left=2mm, right=2mm, title={\footnotesize\textbf{Seasonal temperature change in Atlanta}}]
\footnotesize
\textbf{Pipeline steps} (topological order):\\[1pt]
\begin{tabular}{@{}rl@{\;\;}l@{}}
\texttt{m1--m4} & resolve city $\to$ pixel coords & (metadata, no error) \\
\texttt{m5} & read Jan \& Jul temperature & $T_{\mathrm{Jan}}{=}(5.2, 0, 0, v, 0)$ \\
    &                              & $T_{\mathrm{Jul}}{=}(26.1, 0, 0, v, 0)$ \\
\texttt{m6} & $\Delta T = T_{\mathrm{Jul}} - T_{\mathrm{Jan}}$ & $v_\Delta = 2v$ \\
\texttt{m7} & compose answer & $20.9 \pm 3\sqrt{v_\Delta}$\textdegree C \\
\end{tabular}
\end{tcolorbox}
\caption{An LLM decomposes a climate question into pipeline steps (m1--m7).
Each computation step (m5--m7) maps to a TOPIQ operator; the final answer includes a $99.7\%$ confidence interval derived from compression metadata alone.}
\label{fig:scidag-dag}
\vspace{-1em}
\end{figure}

More complex queries (regional averages, multi-field comparisons, seasonal correlations) decompose into longer chains of the same primitive operators: aggregation with $\alpha$-corrected variance, cross-field multiplication, sigmoid masking.
All operators used in this pipeline correspond to QoI families evaluated in \S\ref{sec:accuracy}, ensuring that the propagation behavior is empirically validated, not merely theoretical.

\paragraph{Comparison with QoI-preserving compression.}
Existing QoI-preserving methods such as QPET~\cite{liu2025vldb} require the QoI and analysis region to be fixed at compression time, which is incompatible with dynamic, query-driven pipelines.
QPET guarantees QoI error bounds on a pre-defined regular grid; however, LLM-driven queries target arbitrary positions and block sizes.
On CESM-ATM, QPET configured for mean$(x^2)$ on a $200{\times}200$ grid leaves up to 11\% of random $100{\times}100$ queries exceeding its grid-level worst case (max exceedance: $4{\times}$ the grid bound), with no uncertainty information for these off-grid results.
In contrast, SZ3 combined with TOPIQ's meta-interpolation achieves $3\sigma$ coverage of $99.9\%$ on the same queries ($\sigma_z = 0.97$, CLDTOT, $\varepsilon_{\mathrm{rel}}{=}10^{-3}$).
The two approaches are complementary: QPET suits fixed pipelines; TOPIQ suits dynamic, exploratory analysis.
\section{Conclusion and Future work}
\label{sec:conclusion}
We presented TOPIQ, a statistical error-propagation framework that predicts QoI-level bias and uncertainty from compact compression metadata by decomposing QoIs into primitive operators with closed-form propagation rules. The framework accounts for spatial error correlation (via the scale-invariant correlation index $\alpha$) and within-field data-error coupling (via $\mathrm{Cov}(x,e)$), achieving well-calibrated predictions in 93.1\% of 552 evaluations across 4 datasets, 3 compressors, 4 QoI families, and 8 error bounds. Pre-computed metadata enables post-hoc uncertainty quantification for arbitrary query regions at 56x--402x speedup over direct computation.

Several directions remain open.
On the operator side, the current second-order Taylor expansion targets smooth functions; truncation-style operators such as ReLU, max, and min lack useful second derivatives and would require fundamentally different propagation techniques, such as piecewise local rules near thresholds or enriching the propagated summaries with distribution-shape information (e.g., skewness and kurtosis).
Discrete QoIs such as feature counts (threshold crossings, pattern counts) also fall outside the smooth-operator scope; however, many such QoIs can be analyzed through a margin-based view, where a count changes only when compression error moves an element across the relevant threshold.
Looking further ahead, integrating the propagation rules into the compression loop could enable QoI-aware error-bound selection: given a target QoI uncertainty budget, the framework could recommend the error bound that satisfies it, bridging the gap between QoI prediction and QoI preservation.

\section*{Acknowledgment}
Argonne National Laboratory’s contribution is based upon work supported by Laboratory Directed Research and Development (LDRD) funding from Argonne National Laboratory, Advanced Scientific Computing Research (ASCR), provided by the Director, Office of Science, of the U.S. Department of Energy under Contract No. DE-AC02-06CH11357.

An award of computer time was provided by the U.S. Department of Energy’s (DOE) Innovative and Novel Computational Impact on Theory and Experiment (INCITE) Program. This research used resources from the Argonne Leadership Computing Facility, a U.S. DOE Office of Science user facility at Argonne National Laboratory, which is supported by the Office of Science of the U.S. DOE under Contract No. DE-AC02-06CH11357.

This work was also supported by the National Science Foundation under Grants OAC-2311875, OAC-2514036, and OAC-2513768. 

The authors used Claude (Anthropic) and Chatgpt (OpenAI) to assist with manuscript editing and
artifact-script development; all technical content, experiments,
and conclusions are the authors' own.





\bibliographystyle{IEEEbib}
\bibliography{refs}

\end{document}